\documentclass{Interspeech}

\interspeechcameraready

\usepackage{colortbl}
\usepackage{amsmath, amssymb}
\definecolor{lightblue}{RGB}{220,248,255}
\usepackage{adjustbox}
\usepackage{amssymb}
\usepackage{marvosym}

\title{Generalizable Audio Deepfake Detection via Hierarchical Structure Learning and Feature Whitening in Poincar\'e sphere}

\author[affiliation={1}]{Mingru}{Yang*}
\author[affiliation={2}]{Yanmei}{Gu*}
\author[affiliation={1}]{Qianhua}{He $^{\dagger}$}
\author[affiliation={1}]{Yanxiong}{Li $^{\dagger}$}
\author[affiliation={1}]{Peirong}{Zhang}
\author[affiliation={1}]{Yongqiang}{Chen}
\author[affiliation={2}]{Zhiming}{Wang}
\author[affiliation={2}]{Huijia}{Zhu $^{\dagger}$}
\author[affiliation={2}]{Jian}{Liu}
\author[affiliation={2}]{Weiqiang}{Wang}

\affiliation{School of Electronic and Information Engineering}{South China University of Technology}{China}
\affiliation{}{Ant Group}{China}
\email{eemryang@mail.scut.edu.cn, eeqhhe@scut.edu.cn, eeyxli@scut.edu.cn}

\usepackage{comment}

\begin{document}

\maketitle

\begin{abstract}

Audio deepfake detection (ADD) faces critical generalization challenges due to diverse real-world spoofing attacks and domain variations. However, existing methods primarily rely on Euclidean distances, failing to adequately capture the intrinsic hierarchical structures associated with attack categories and domain factors. To address these issues, we design a novel framework \textbf{Poin-HierNet} to construct domain-invariant hierarchical representations in the Poincaré sphere. Poin-HierNet includes three key components: 1) Poincaré Prototype Learning (PPL) with several data prototypes aligning sample features and capturing multilevel hierarchies beyond human labels; 2) Hierarchical Structure Learning (HSL) leverages top prototypes to establish a tree-like hierarchical structure from data prototypes; and 3) Poincaré Feature Whitening (PFW) enhances domain invariance by applying feature whitening to suppress domain-sensitive features. We evaluate our approach on four datasets: ASVspoof 2019 LA, ASVspoof 2021 LA, ASVspoof 2021 DF, and In-The-Wild. Experimental results demonstrate that Poin-HierNet exceeds state-of-the-art methods in Equal Error Rate.  


\end{abstract}

\section{Introduction}

Audio deepfake detection (ADD) is critical for protecting speaker verification systems from various attacks, including speech synthesis, voice conversion and voice cloning. Significant progress has been made in developing both front-end features \cite{2022wav2vec-aasist,guo2024wavlm-mfa,zhang2024xls-r} and back-end models \cite{jung2022aasist,chen2024rawbmamba,wu2024adapter}, achieving promising results in intra-database deepfake detection. However, generalizing to diverse unseen domains remains challenging, particularly in real-world scenarios with diverse, unpredictable attacks and spoofed utterances spanning various domains like channels, codecs, and compression formats \cite{asvspoof2021,in-the-wild-dataset}. 

\begin{figure}[t]
\centering
\includegraphics[width=0.5\textwidth]{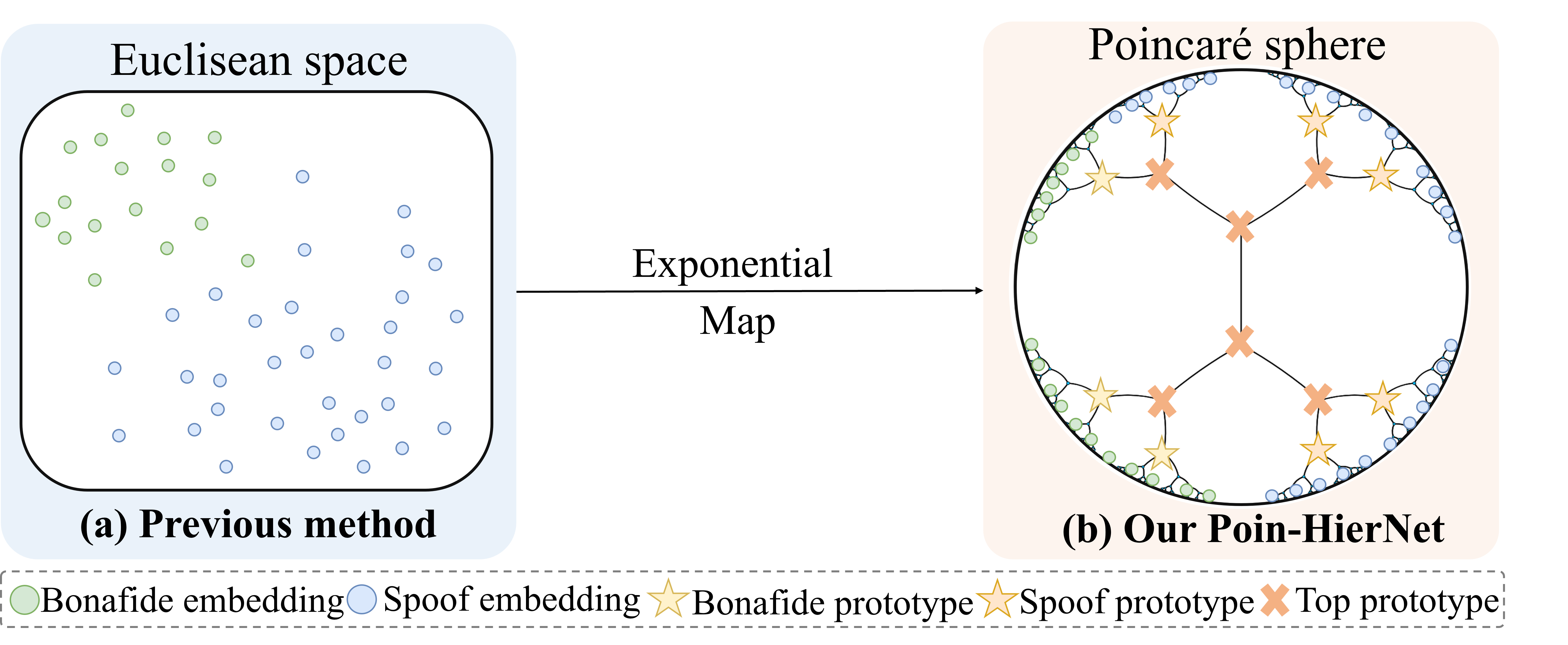}
\caption{Motivation of Poin-HierNet. Previous works represent data distribution in Euclidean space only based on human-labeled classes, whereas Poin-HierNet constructs hierarchical feature space in the Poincaré ball model.}
\label{motivation}
\end{figure}

To develop generalizable deepfake detectors, existing methods focus on two directions: 1) refining learning strategies to capture complex forgery patterns, and 2) using data augmentation to simulate unseen attacks. The first includes domain adversarial learning \cite{xie2023domain-adversarial} and knowledge distillation \cite{wang2024ftdkd-distillation,ren2023lightweight} for domain-invariant feature extraction, one-class learning \cite{kim2024one,wang2024genuine-gfl-fad} for bonafide audio alignment, contrastive learning \cite{tran2024spoofed-contrastive,doan2024balance} for spoof-genuine discrimination, and prototype-based methods \cite{huang2025LDA_LA} for feature space modeling. Data augmentation methods introduce perturbations, including speed perturbation, SpecAugment \cite{park2019specaugment}, and RawBoost \cite{tak2022rawboost}. However, as shown in Fig.~\ref{motivation}(a), all these methods operate in Euclidean space and rely on human-labeled classes, neglecting the intrinsic hierarchy of 
\vfill
\noindent
\begin{minipage}[t]{0.45\textwidth}
    \rule{2.9cm}{0.3pt} \\[-0.5em]  
    \scriptsize
     \\
     * Equal Contribution.\\
     † Corresponding Authors.
\end{minipage}
ADD data.  
This hierarchical structure is evident because attack algorithms form distinct categories, whereas domain factors (e.g., channels, codecs, compression methods) define finer subcategories. Leveraging this structure is key to understanding the data and improving model generalization against unseen attacks, yet it remains unexplored in audio deepfake detection. 

In this paper,
we introduce the Poincaré ball model, which is inherently well-suited for representing hierarchical data structures \cite{nickel2017poincare,kim2023hier}. Then we propose \textbf{Poin-HierNet} (Fig.~\ref{motivation}(b)), a framework that constructs a domain-invariant hierarchical feature space in the Poincar\'e sphere. Poin-HierNet comprises three key learning processes: Poincar\'e Prototype Learning (PPL), Hierarchical Structure Learning (HSL), and Poincar\'e Feature Whitening (PFW). 
PPL introduces a set of learnable data prototypes to represent the overall feature distribution, aligning these prototypes with sample features using the constraint $\mathcal{L}_{PPL}$. Multiple prototypes are assigned to each bonafide and spoof class, thereby capturing hierarchical structures beyond human-labeled categories. 
HSL integrates the top prototypes and leverages the Poincaré ball model to explore the correlations between data prototypes and top prototypes, thus establishing a tree-like hierarchical structure. 
PFW applies feature whitening to suppress domain-sensitive features, ensuring the learned hierarchical feature space remains domain-invariant. 


\begin{figure*}[t]
    \centering
    \includegraphics[width=0.91\textwidth]{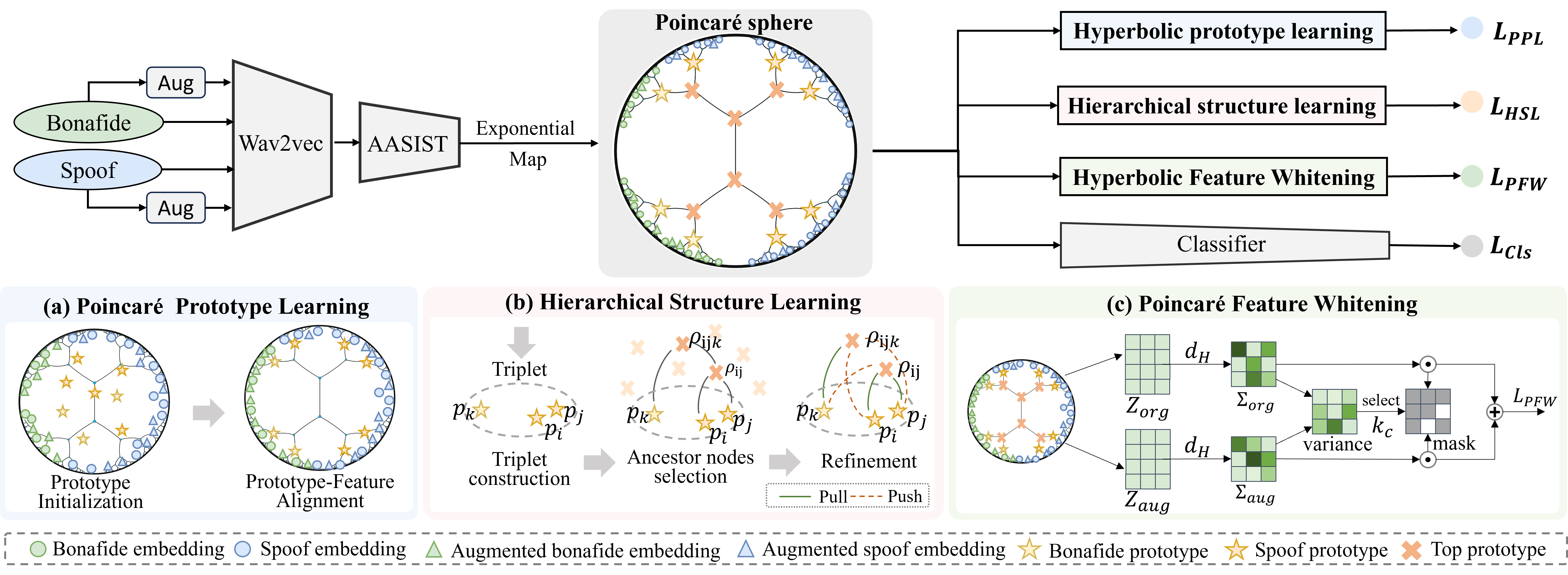}
    \caption{Overall framework of the proposed Poin-HierNet. \textbf{TOP}: Illustration of the training procedures for Poin-HierNet. \textbf{Bottom}: Schematic of Poincar\'e Prototype Learning (\textbf{a}), Hierarchical Structure Learning (\textbf{b}), and Poincar\'e Feature Whitening (\textbf{c}) components.}
    \label{framework}
\end{figure*}

In summary, our main contributions are as follows. We propose Poin-HierNet, a novel framework that constructs a domain-invariant hierarchical feature space. To the best of our knowledge, this is the first method that addresses generalizable audio deepfake detection from the perspective of hierarchical space. The Poin-HierNet consists of PPL, HSL, and PFW. PPL and HSL model the data distribution and establish a hierarchical structure through prototype learning, while PFW applies feature whitening to ensure that this hierarchical structure remains domain-invariant. Experimental results on multiple datasets verify that our method outperforms state-of-the-art methods in terms of Equal Error Rate (EER).





\section{Method}
The overall architecture of the proposed Poin-HierNet is depicted in Fig.~\ref{framework}. Following prior studies \cite{2022wav2vec-aasist,huang2025LDA_LA}, we utilize the wav2vec 2.0 XLS-R (0.3B) \cite{babu2021xls-r} as the frontend feature extractor and AASIST \cite{jung2022aasist} as the backend model. Bonafide and spoof audio samples are processed by wav2vec 2.0 XLS-R and AASIST, obtaining high-dimensional features in Euclidean space. We use paired inputs, specifically the original sample and its augmentation, both processed identically. These features are then mapped onto the Poincaré sphere via exponential mapping. To construct domain-invariant hierarchical representations in the Poincaré sphere, PPL represents the overall data distribution through prototype learning by enhancing prototype-feature and original-augmentation alignments; HSL introduces triplet construction to establish the hierarchical structure; and PFW calculates a mask matrix to suppress domain-sensitive features, ensuring that the hierarchical feature space remains domain-invariant. Finally, the overall loss is integrated for optimization.

\subsection{The Poincar\'e ball model}
Let $\mathbb{E}^n$ and $\mathbb{H}^n$ denote the n-dimensional Euclidean space and hyperbolic space, respectively. The Poincar\'e ball model $(\mathbb{H}_c^n,g^\mathbb{H})$ is given by the manifold:
\begin{equation}
\mathbb{H}_c^n = \{\mathbf{x} \in \mathbb{E}^n:c\lVert \mathbf{x} \rVert<1\},
\end{equation}
equipped with the Riemannian metric: 
\begin{equation}\label{2}
    g^\mathbb{H} = \left(\frac{2}{1-c\lVert \mathbf{x}\rVert^2}\right)^2 g^\mathbb{E},
\end{equation}
where c is the curvature hyperparameter, $\lVert \cdot \rVert$ denotes the Euclidean norm, $g^\mathbb{E} = \mathbf{I}_n$ is the Euclidean metric.

The mapping function converting Euclidean embeddings to Poincaré embeddings via the exponential map is defined as:
\begin{equation}
\exp_\mathbf{0}^c(\mathbf{v}) = \tanh\sqrt{c}\lVert \mathbf{v}\rVert\frac{\mathbf{v}}{\sqrt{c}\lVert \mathbf{v}\rVert}. 
\end{equation}
The distance between points $\mathbf{u}, \mathbf{v} \in \mathbb{H}^n$ is given as:
\begin{equation}\label{dh}
d_H(\mathbf{u},\mathbf{v})=arcosh\left(1+2\frac{\lVert \mathbf{u}-\mathbf{v} \rVert^2}{(1-\lVert \mathbf{u}\rVert^2)(1-\lVert \mathbf{v}\rVert^2)} \right).
\end{equation}


\subsection{Poincar\'e prototype learning (PPL)}
To facilitate the construction of a hierarchical structure within the Poincaré sphere, we introduce prototype learning. Distinct learnable prototype sets are employed for each category to represent their essential features. Specifically, $K_b$ bonafide prototypes and $K_s$ spoofing prototypes are established to represent the typical characteristics of genuine and manipulated features, respectively. 
These trainable prototypes are randomly initialized in the Poincaré sphere and formally defined as  $P = \{p_j^r \in \mathbb{R}^{D}\}$, where $j \in \{0,1\}$ denotes the class type (with 0 representing bonafide samples and 1 representing spoof samples). The index $r$ ranges from $1$ to $K_j$, where $K_j$ corresponds to the number of prototypes for each class (i.e.,$K_j = K_b$ for bonafide samples when $j = 0$, and $K_j = K_s$ for spoof samples when $j = 1$). Here, $D$ represents the dimension of the feature space. 



Inspired by \cite{prototype-learning, hu2024rethinking}, we introduce the following loss for Poincar\'e prototype learning:  

\begin{align}
\mathcal{L}_{proto} = -\sum_{n=1}^{N} \log \frac{\exp(-d_H(z_n, p_{y_n}^{k}))}
{\sum_{r=1}^{K_j} \exp(-d_H(z_n, p_j^r))},
\end{align}
where $d_H$ is the hyperbolic distance defined in Eq.(\ref{dh}),  $z_n \in \mathbb{R}^{D}$ represents the embedding of the $n$-th sample, $y_n$ denotes its label, and $p_{y_n}^k$ is the prototype associated with the $n$-th sample, i.e., the closest prototype from $\{p_{y_n}^{r}\}$ to that sample within the Poincaré ball model. As shown in Fig.~\ref{framework}(a), this constraint brings the features close together with their corresponding prototypes while driving them far apart from non-corresponding prototypes.

Moreover, to ensure that the learned prototypes effectively generalize to unknown encoding and transmission conditions, 
we implement the following loss function to enforce the augmented feature $z^{aug} \in \mathbb{R}^{D}$ aligns with the original feature $z$: 


\begin{equation}
\begin{tiny}
\mathcal{L}_{aug} = \sum_{n=1}^{N} \left( d_H(z_n, z_n^{aug}) + \left\lVert d_H(z_n, p_{y_n}^{k}) - d_H(z_n^{aug}, p_{y_n}^{k}) \right\rVert_2 \right).
\end{tiny}
\end{equation}

This loss minimizes the distance between $z_{n}$ and $z_{n}^{aug}$ as well as the discrepancy between their distances to the corresponding prototype, thus learning task-specific features and enhancing Poincaré prototype representations. The total loss for Poincar\'e prototype learning is: 
\begin{align}
\mathcal{L}_{PPL} = \mathcal{L}_{proto} + \mathcal{L}_{aug}.
\end{align}

\subsection{Hierarchical structure learning (HSL)}



Based on the data prototypes learned in Poincaré sphere that capture the sample distribution, we subsequently explore various latent semantic hierarchies within the Poincar\'e prototypes by modeling a tree-like hierarchical structure. 
We draw inspiration from \cite{kim2023hier} and employ a series of learnable top prototypes, $P_{top}=\{ \rho_k| k \in \{0,1,\ldots, K_{top}\}\}$, as higher-level nodes. As shown in Fig.~\ref{framework}(b), HSL consists of three stages: triplet construction, ancestor nodes selection, and hierarchical distance refinement.

\textbf{Triplet construction.} To capture structural information beyond labels, we omit label details and flatten the learned Poincar\'e prototypes as $P = \{p_n|n \in \{0,1,\ldots,(K_b+K_s)\}\}$. Subsequently, we randomly select a prototype as the anchor $p_i$. 
A positive pair is constructed by choosing a random sample $p_j$ from its $K$-nearest neighbors $R_K(p_i)$, while a negative pair is formed by randomly selecting a prototype $p_k$ from outside the neighborhood $R_K(p_i)$.
The triplets are expressed as: 
\begin{align}
T = \{(p_i,p_j,p_k)|(p_j \in R_K(p_i)) \wedge (p_k \notin R_K(p_i))\}.
\end{align}

\textbf{Ancestor nodes selection.} Higher-level nodes are identified within the top-level prototypes $P_{top}$. As illustrated in Fig.~\ref{framework}(b), in the triplet, $p_i$ and $p_j$ encouraged to share the same lowest common ancestor (LCA) $\rho_{ij}$, while $p_i$ and $p_k$ have different LCAs. The selection of $\rho_{ij}$ from $P_{top}$ follows:
\begin{align}\label{top-node-selection}
\rho_{ij}=\arg\max_{\rho}(e^{(-max\{d_H(p_i,\rho), d_H(p_j,\rho)\})}+g_{ij}),
\end{align}
where $e^{(-max\{d_H(p_i,\rho), d_H(p_j,\rho)\})}$ represents the probability of $\rho$ being the ideal representative $\rho_{ij}$, and $g_{ij}$ is a noise term sampled from a Gumbel(0,1) distribution to mitigate the risk of local optima.
The same approach as in Eq. (\ref{top-node-selection}) is used to select the LCA of $\rho_{ij}$ and $\rho_{k}$, denoted as $\rho_{ijk}$, as a higher-level node than $\rho_{ij}$.

\textbf{Hierarchical distance refinement.} The optimization based on the distance relationship between prototypes and their higher-level prototype ancestors is as follows: 
\begin{align}
\mathcal{L}_{HSL} = \left[ d_H(p_i, \rho_{ij}) - d_H(p_i, \rho_{ijk}) + \delta \right]  \nonumber \\
+ \left[ d_H(p_j, \rho_{ij}) - d_H(p_j, \rho_{ijk}) + \delta \right] \\
+ \left[ d_H(p_k, \rho_{ijk}) - d_H(p_k, \rho_{ij}) + \delta \right], \nonumber 
\end{align}
where $\delta$ represents the margin. 

\subsection{Poincar\'e feature whitening (PFW)}
To capture domain-invariant features at a finer granularity in the Poincar\'e sphere, we propose Poincar\'e Feature Whitening (PFW). PFW imposes explicit constraints on the correlations among feature dimensions, suppressing those sensitive to domain variations while accentuating the insensitive ones. 

Prior works \cite{li2017universal,roy2019unsupervised, zhou2023instance} have demonstrated that the feature covariance matrix captures domain-specific features in Euclidean space; however, due to the distinct geometry and metric of Poincar\'e sphere, the conventional covariance matrix cannot be directly computed. To address this issue, we leverage the hyperbolic metric by replacing the covariance matrix with a similarity matrix computed across different feature dimensions. The similarity matrix is defined as:
\begin{align}\label{sim-matrix}
\Sigma = d_H(Z  ,Z^\top) \in \mathbb{R}^{D \times D}, 
\end{align}
where $Z^\top \in \mathbb{R}^{B \times D}$ represents the output features, 
with $B$ indicating the batch size and $D$ denoting the feature dimension. 

As shown in Fig.~\ref{framework}(c), PFW is computed as follows. For the original and augmented features, the similarity matrices $\Sigma_{org}$ and $\Sigma_{aug}$ are computed using eq. (\ref{sim-matrix}), respectively. Next, we derive a mask $M$ to identify positions in the similarity matrix that are sensitive to domain-specific styles. Concretely, given $\Sigma_{org}$ and $\Sigma_{aug}$, we compute the element-wise variance as follows: 
\begin{align}
\mu_{\Sigma} = \frac{1}{2}(\Sigma_{org} + \Sigma_{aug}) \in \mathbb{R}^{D \times D}, \nonumber \\
\sigma_{\Sigma}^2 = \frac{1}{2}((\Sigma_{org} - \mu_{\Sigma})^2+(\Sigma_{aug} - \mu_{\Sigma})^2) \in \mathbb{R}^{D \times D}.
\end{align}
We then sort all the elements of $\sigma_{\Sigma}^2$, and select the top $k_c$ fraction, setting them to 1 to obtain the mask $M \in \mathbb{R}^{D \times D}$: 
\begin{align}
M_{i,j}(k_c) = 
\begin{cases} 
1, & \text{if } \text{index}(\sigma^2_{\Sigma}(i,j)) < \text{len}(\sigma^2_{\Sigma}) \times k_c \\ 
0, & \text{otherwise}
\end{cases}
\end{align}
where the selection ratio $k_c=k_b$ for bonafide utterances and $k_c=k_s$ for spoof ones.

Finally, we adopt this mask to perform PFW, which forces the selected features to be suppressed:
\begin{align}
\mathcal{L}_{\text{PFW}} = \sum_{k_c \in \{k_b, k_s\}} \sum_{t \in \{org, aug\}} E \left[ \lVert  \Sigma_t \odot M(k_c) \rVert \right],
\end{align}
where $E$ is the arithmetic mean.

\subsection{Overall training and optimization}
Building on the previous learning process, Poincar\'e prototypes precisely locate instance features. Consequently, Poin-HierNet predicts the final task label by classifying the similarity between sample features $z_n$ and prototypes $P$, using Binary Cross Entropy (BCE) as the prediction loss:  
\begin{align}
\mathcal{L}_{cls} = \mathrm{BCE}(W \cdot d_H(z_n,P),y_n),
\end{align}
where $y_n$ is the label for $z_n$. 
The overall training loss is:  
\begin{align}
\mathcal{L}_{all} = \mathcal{L}_{cls} + \mathcal{L}_{PPL} + \mathcal{L}_{HSL} + \mathcal{L}_{PFW}.
\end{align}

During the evaluation phase, only the original samples $x$ and $\mathcal{L}_{cls}$ are used. 

\section{Experiments}
\subsection{Datasets and metrics}

The ASVspoof 2019 LA dataset \cite{asvspoof2019} is utilized for training, comprising a total of 25k utterances, which include both bonafide and spoofed samples. The spoofed samples are generated through six different attack types, including voice conversion and speech synthesis techniques. To assess generalization performance, four test datasets from diverse domains are employed: the ASVspoof 2019 LA evaluation set (19LA) \cite{asvspoof2019}, which includes 71k utterances with 13 distinct spoofing attacks; the ASVspoof 2021 LA set (21LA) \cite{asvspoof2021}, containing 181k utterances with methods similar to those in 19LA, while also considering encoding and transmission effects from telephony systems; the ASVspoof 2021 DF set (21DF) \cite{asvspoof2021}, with over 600k
utterances and more than 100 spoofing attacks processed with various lossy codecs; and the In-The-Wild dataset (ITW) \cite{in-the-wild-dataset}, which includes 32k utterances collected under real-world, non-controlled conditions, making it a more challenging dataset. Performance is measured with EER.

\subsection{Implementation details}

All input utterances are randomly chunked into 4-second segments, with the fifth algorithm of the Rawboost \cite{tak2022rawboost} applied as the basic augmentation. These samples $x$ are further augmented by adding stationary, signal-independent additive noise \cite{tak2022rawboost} to obtain the augmented samples $x_{aug}$. 
For the ASVspoof 2021 LA evaluation set, we additionally incorporate noises from the RIR corpus \cite{snyder2015musan}, and apply the codec augmentation for better performance. The codec types include adts, mp3, ogg, a-law, and $\mu$-law \cite{asvspoof2021}. 

The Adam optimizer is utilized, with a learning rate of 1e-6 for the backbone model and 1e-3 for the prototypes. The feature dimension in the Poincar\'e sphere is set to 160, with a curvature parameter $c$ of 0.01. The number of bonafide prototypes $K_b$ and spoofing prototypes $K_s$ is set to 10 and 6, respectively. For $\mathcal{L}_{HSL}$, we select three nearest neighbors to construct positive prototype pairs. The number of top prototypes $K_{top}$ is set to 256, with the margin $\delta$ established at 0.1. For $\mathcal{L}_{PFW}$, the values of $k_b$ and $k_s$ are empirically set at 0.3\% and 0.06\%,respectively. The batch size $B$ is 256, with the bonafide batch size defined as $B * K_b/K_s$ and the spoofing batch size calculated as $B-(B * K_b/K_s)$. This approach ensures balanced learning of the data distribution during the Poincaré prototype learning process.

\subsection{Comparison with existing methods}
We evaluate the performance of Poin-HierNet against state-of-the-art (SOTA) methods in terms of the EER. 
These baselines consist of a variety of network architectures and domain generalization techniques. All methods, including Poin-HierNet, are trained on the ASVspoof2019 LA training set and assessed on four datasets. The results are presented in Table~\ref{tab:comparison}.

\begin{table}[h]
\centering
\caption{Comparison of EER (\%) performance for different methods across multiple datasets. All systems are trained on the ASVspoof2019 LA training set.}
\label{tab:comparison}
\begin{tabular}{l|cccc}
\toprule
System & 19LA & 21LA & 21DF & ITW \\
\midrule
WavLM+AttM \cite{2024wavlm-attm} & 0.65 & 3.50 & 3.19 & - \\
Wav2Vec+LogReg \cite{2024wav2vec-logreg} & 0.50 & - & - & 7.20 \\
WavLM+MFA \cite{guo2024wavlm-mfa} & 0.42 & 5.08 & 2.56 & - \\
FTDKD \cite{wang2024ftdkd-distillation} & 0.30 & 2.96 & 2.82 & - \\
OCKD \cite{lu2024ockd} & 0.39 & \underline{0.90} & 2.27 & 7.68 \\
GFL-FAD \cite{wang2024genuine-gfl-fad} & 0.25 & - & - & - \\
WavLM+ASP \cite{tran2024spoofed-contrastive} & 0.23 & 3.31 & 4.47 & - \\
Wav2Vec+Linear \cite{2023wav2vec-linear} & 0.22 & 3.63 & 3.65 & 16.17 \\
OC+ACS \cite{kim2024one} & 0.17 & 1.30 & \underline{2.19} & - \\
Wav2Vec+AASIST \cite{2022wav2vec-aasist} & - & \textbf{0.82} & 2.85 & - \\
Wav2Vec+AASIST2 \cite{2024wav2vec-aasist2} & \underline{0.15} & 1.61 & 2.77 & - \\
LSR+LSA \cite{huang2025LDA_LA} & \underline{0.15} & 1.19 & 2.43 & \underline{5.92} \\
\rowcolor{lightblue}
\textbf{Poin-HierNet(Ours)} & \textbf{0.11} & 0.94 & \textbf{1.40} & \textbf{4.91} \\
\bottomrule
\end{tabular}
\end{table}

As presented in Table~\ref{tab:comparison}, Poin-HierNet outperforms all existing methods, achieving SOTA results on the 19LA, 21DF, and ITW datasets with EERs of 0.11\%, 1.40\%, and 4.91\%, respectively. Notably, the 21DF and ITW datasets present significant challenges, and the substantial reduction in EER achieved on these datasets further illustrates the robust generalization capabilities of our approach.
On the 21LA dataset, it also achieves a competitive EER of 0.94\%. Unlike existing baselines that learn data representations in Euclidean space, Poin-HierNet models domain-invariant features in the Poincaré sphere and builds a hierarchical structure upon them.
This fundamental difference contributes to the superior performance of Poin-HierNet.


\subsection{Ablation study}
This section conducts ablation studies to assess the impact of each loss function and the number of prototypes on the performance of the proposed method.

\begin{table}[h]
\centering
\caption{Ablation study on the loss functions across multiple datasets in terms of EER(\%). The checked items (\checkmark) indicate the corresponding loss is used for optimization.}
\label{tab:ablation-loss}
\begin{tabular}{ccc|cccc}
\toprule
$\mathcal{L}_{PPL}$ & $\mathcal{L}_{HSL}$ & $\mathcal{L}_{PFW}$ & 19LA & 21LA & 21DF & ITW \\
\midrule
\checkmark &  &  & 0.24 & 1.45 & 1.95 & 5.93 \\
\checkmark & \checkmark &   & 0.16  & 1.30  & 1.88  & 5.56 \\
\checkmark & \checkmark & \checkmark & 0.11 & 0.94 & 1.40 & 4.91 \\
\bottomrule
\end{tabular}
\end{table}

As shown in Table~\ref{tab:ablation-loss}, we present the results to evaluate the effects of different loss functions by progressively adding the three loss functions across multiple datasets.
The results demonstrate that using only $\mathcal{L}_{PPL}$ achieves competitive performance, as PPL effectively aligns prototypes with feature distributions in the Poincaré space while minimizing information loss during alignment.
After incorporating $\mathcal{L}_{HSL}$, performance improves significantly, as hyperbolic structure learning leverages the advantages of the Poincaré sphere to establish a more meaningful hierarchical feature space. 
Finally, adding $\mathcal{L}_{PFW}$ yields the best performance on all four datasets, since PFW  compels the model to focus on learning domain-invariant features.

\begin{table}[h]
\centering
\caption{Ablation study on the number of bonafide prototypes $K_b$ and spoofing prototypes $K_s$ across multiple datasets in terms of EER(\%).}
\label{tab:ablation-prototype}
\begin{tabular}{cc|cccc}
\toprule
$K_{b}$ & $K_{s}$ & 19LA & 21LA & 21DF & ITW \\
\midrule
 12 & 4 & 0.14  & 1.55  & 1.83 & 5.03 \\
 10 & 6 & 0.11 & 0.94 & 1.40 & 4.91 \\
 8 & 8 & 0.17 & 2.13 & 2.15 & 5.65 \\
 6 & 10 & 0.21 & 2.96 & 2.02 & 5.82 \\
 4 & 12 & 0.31 & 2.92 & 2.44 & 7.96 \\
\bottomrule
\end{tabular}
\end{table}


Table~\ref{tab:ablation-prototype} presents the influence of the values of $K_b$ and $K_s$ on the performance of our method. As observed, the best performance is achieved with $K_b=10$ and $K_s=6$. Excessive spoofing prototypes introduce redundancy and degrade performance, while an insufficient number of bonafide prototypes weakens the model’s generalization capability. These findings highlight the importance of prototype selection in hyperbolic representation learning for anti-spoofing tasks.


\section{Conclusions}
In this paper, we propose Poin-HierNet, a novel framework for constructing a domain-invariant hierarchical feature space using the Poincaré ball model. 
Within Poin-HierNet, PPL introduces a set of data prototypes to represent the overall data distribution; HSL explores the correlations between data prototypes and top prototypes to establish a tree-like hierarchical structure; and PFW applies feature whitening to enhance the model's ability to achieve domain invariance. Experimental results demonstrate the superiority of our framework over existing methods. 


\section{Acknowledgments}

This work was partially supported by the Ant Group Research Intern Program and the Guangdong Science and Technology Foundation (2023A0505050116), as well as by Guangdong Natural Science Foundation (2022A1515011687), the National Natural Science Foundation of China (62371195), the National Key R\&D Program of China (2022YFC3301703), and the Pioneer R\&D Program of Zhejiang Province (No. 2024C01024).


\bibliographystyle{IEEEtran}
\bibliography{mybib}

\end{document}